\documentclass[11pt, logo]{googledeepmind}
\usepackage[authoryear, sort&compress, round]{natbib}

\usepackage[inkscapelatex=false]{svg}

\usepackage[most, breakable, skins]{tcolorbox}
\tcbuselibrary{skins}
\usepackage{lipsum}
\usepackage{tabularx}
\usepackage{subcaption}
\usepackage{authblk}
\usepackage{multirow}
\usepackage{multicol}
\usepackage{array}
\usepackage{listings, listings-rust}
\usepackage{fontawesome5}
\usepackage{pgfplots}
\usepackage{amssymb,graphicx}
\usepackage[dvipsnames]{xcolor}
\usepackage{hyperref}

\newcolumntype{L}[1]{>{\raggedright\let\newline\\\arraybackslash\hspace{0pt}}m{#1}}
\newcolumntype{C}[1]{>{\centering}m{#1}}

\newcolumntype{R}[1]{>{\raggedleft\let\newline\\\arraybackslash\hspace{0pt}}m{#1}}

\definecolor{ao}{rgb}{0.0, 0.0, 1.0}

\newcommand\vcent[1]{\vcenter{\hbox{#1}}}
\newcommand\loudspeaker[1][3]{\ensuremath{\vcent{\rule{.6ex}{.6ex}}\kern-.5ex%
  \vcent{\scalebox{.6}[1]{\rotatebox[origin=center]{90}{$\blacktriangle$}}}%
  \ifnum#1>0\relax\kern.05ex\vcent{\scalebox{.4}{\ttfamily)}}%
  \ifnum#1>1\relax\kern-.4ex\vcent{\scalebox{.56}{\ttfamily)}}%
  \ifnum#1>2\relax\kern-.55ex\vcent{\scalebox{.7}{\ttfamily)}}%
  \fi\fi\fi}%
}
\renewcommand\Affilfont{\itshape\small}
\makeatletter
\renewcommand\AB@affilsepx{\\[\affilsep] \protect\Affilfont}
\let\cite\citep

\title{GeoGPT-RAG Technical Report}

\reportnumber{} 

\renewcommand{\thefootnote}{}

\author[1]{Fei Huang}
\author[1]{Fan Wu}
\author[1]{Zeqing Zhang}
\author[1]{Qihao Wang}
\author{Long Zhang}
\author{Grant Michael Boquet}
\author{Hongyang Chen}

\affil[1]{GeoGPT Team\quad Zhejiang Lab}
\begin{abstract}
GeoGPT is an open large language model system built to advance research in the geosciences. To enhance its domain-specific capabilities, we integrated Retrieval-Augmented Generation (RAG), which augments model outputs with relevant information retrieved from an external knowledge source. GeoGPT uses RAG to draw from the GeoGPT Library, a specialized corpus curated for geoscientific content, enabling it to generate accurate, context-specific answers. Users can also create personalized knowledge bases by uploading their own publication lists, allowing GeoGPT to retrieve and respond using user-provided materials. To further improve retrieval quality and domain alignment, we fine-tuned both the embedding model and a ranking model that scores retrieved passages by relevance to the query. These enhancements optimize RAG for geoscience applications and significantly improve the system’s ability to deliver precise and trustworthy outputs.

GeoGPT reflects a strong commitment to open science through its emphasis on collaboration, transparency, and community-driven development. As part of this commitment, we have open-sourced two core RAG components—\href{https://huggingface.co/GeoGPT-Research-Project/Llama3.1-70B-GeoGPT}{GeoEmbedding} and \href{https://huggingface.co/GeoGPT-Research-Project/Qwen2.5-72B-GeoGPT}{GeoReranker}—to support geoscientists, researchers, and professionals worldwide with powerful, accessible AI tools.
\end{abstract}

\pgfplotsset{compat=1.18} 
\begin{document}
\maketitle
\renewcommand{\thefootnote}{\arabic{footnote}}
\begin{center}

\faGithub\ \href{https://github.com/GeoGPT-Research-Project/GeoGPT-RAG}{https://github.com/GeoGPT-Research-Project/GeoGPT-RAG}

\raisebox{-0.2em}{\includegraphics[height=1.1em]{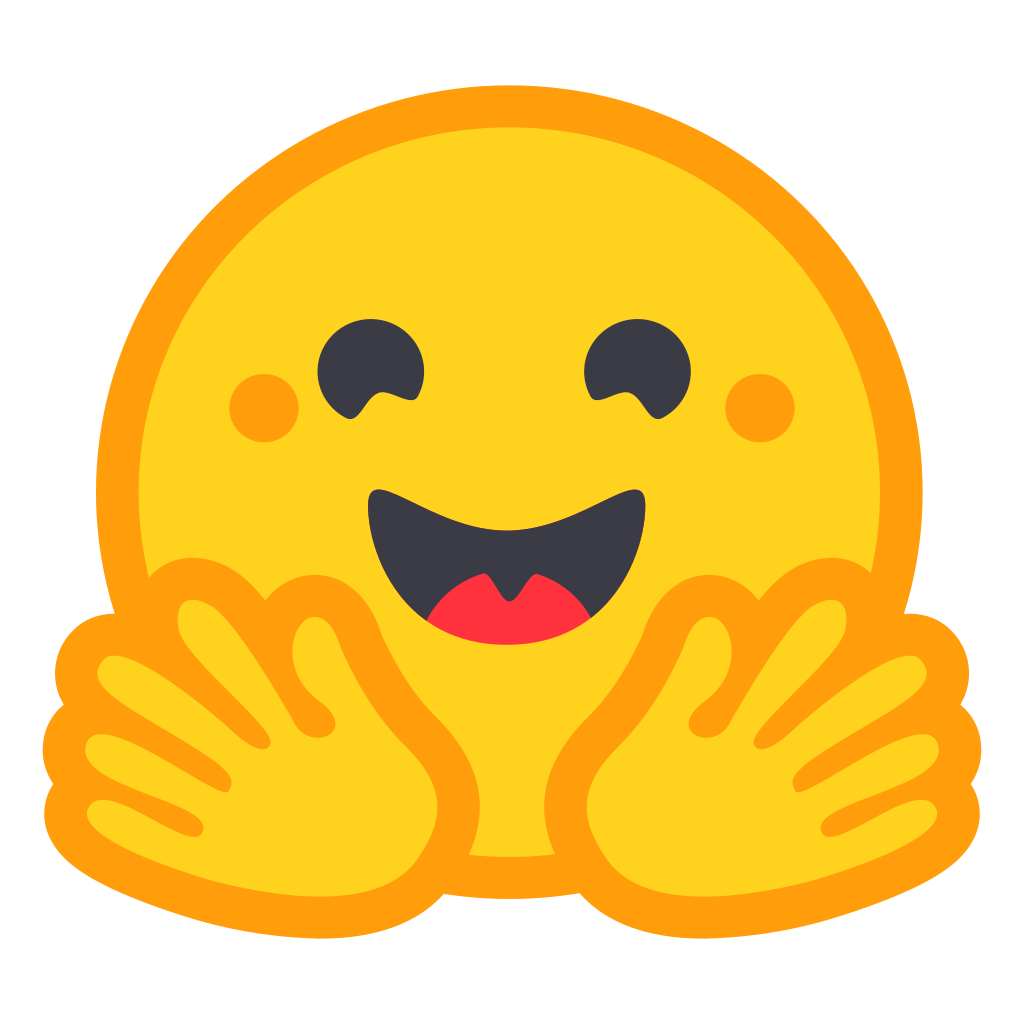}}\ \href{https://huggingface.co/GeoGPT-Research-Project/GeoEmbedding}{https://huggingface.co/GeoGPT-Research-Project/GeoEmbedding}

\raisebox{-0.2em}{\includegraphics[height=1.1em]{figs/huggingface-color.png}}\ \href{https://huggingface.co/GeoGPT-Research-Project/GeoReranker}{https://huggingface.co/GeoGPT-Research-Project/GeoReranker}

\raisebox{-0.2em}{\includegraphics[height=1.1em]{figs/huggingface-color.png}}\ \href{https://huggingface.co/datasets/GeoGPT-Research-Project/GeoRAG-QA}{https://huggingface.co/datasets/GeoGPT-Research-Project/GeoRAG-QA}

\end{center}






\label{sec:application}

\section{Introduction}

\noindent Large language models (LLMs) have demonstrated impressive general-purpose capabilities, but they often fall short in specialized domains due to their tendency to generate factually incorrect or unverifiable information. Retrieval-Augmented Generation (RAG) addresses this limitation by grounding model outputs in external knowledge sources. In fields like geoscience, where precision and interpretability are essential, RAG enables models to produce context-aware and trustworthy responses. It also allows users to upload their own document collections, enabling retrieval from personalized knowledge bases tailored to specific research needs.

\noindent Beyond improving accuracy, RAG supports dynamic updates to its knowledge base, ensuring that outputs reflect the most current information available. It also promotes transparency by making the retrieval and generation process interpretable and traceable—critical for building user confidence in scientific and high-stakes applications.

\noindent RAG enhances LLM performance by retrieving semantically relevant content from external sources using similarity-based search methods. By referencing this material during generation, RAG significantly reduces hallucinations and improves factual consistency. Its integration into modern LLM pipelines has made it a foundational technique for building robust AI systems, particularly in applications like chatbots, research assistants, and domain-specific tools.

\noindent Initially proposed in 2020~\cite{lewis2020retrieval}, RAG has seen rapid adoption since 2023 as its practical advantages over traditional fine-tuning have become evident. These include lower computational costs, improved interpretability, and easier adaptation to new or evolving knowledge~\cite{gao2023retrieval}.

\noindent At its core, RAG combines information retrieval with text generation in a unified workflow. It coordinates multiple components—including document retrieval, reading comprehension, response synthesis, and post-processing—into a flexible architecture. To support a range of use cases, we implemented both a lightweight (native) RAG pipeline and more advanced configurations with varying levels of complexity.

\begin{figure}[h!]
\centering
 \includegraphics[width=0.9\textwidth]{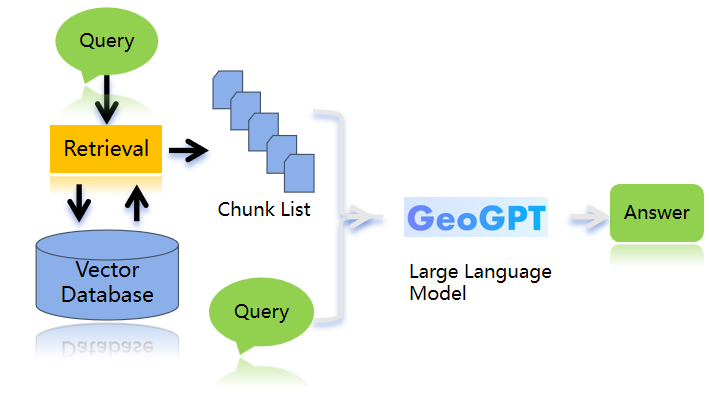}
 \caption{Retrieval-Augmented Generation for GeoGPT}
 \label{fig6-1-1-0}
\end{figure}

\noindent In practice, RAG retrieves relevant information and integrates it directly into the model’s prompt. This augmentation allows the model to ground its responses in verifiable context, resulting in outputs that are not only more accurate but also more transparent and easier to trust.

\noindent \textbf{Main Contributions.} Our work makes the following key contributions to the field of geographic information retrieval with large language models:
\begin{itemize}

    \item \textbf{GeoEmbedding: the first decoder-based embedding model optimized for geoscience.}  
    We introduce \textit{GeoEmbedding}, the first text embedding model built on a modern decoder-only architecture (Mistral-7B) and specifically adapted to the geoscience domain. By incorporating synthetic domain-specific data during training, GeoEmbedding achieves significantly improved retrieval accuracy on geoscience queries while preserving strong general-purpose performance.

    \item \textbf{GeoReranker: the first reranking model tuned for geoscience relevance.}  
    We propose and release \textit{GeoReranker}, a domain-augmented reranking model that improves ranking precision in geoscience applications. Despite its specialization, GeoReranker remains broadly applicable, retaining high performance on general information retrieval tasks.

    \item \textbf{LLM-based synthetic data generation pipeline for geoscience retrieval.}  
    To address the lack of high-quality labeled data in the geoscience domain, we design a novel synthetic data generation pipeline. This pipeline produces realistic and diverse training data by abstracting and reformatting scholarly geoscience articles, enabling robust training of retrieval and reranking models under data-scarce conditions.

    \item \textbf{Data quality enhancement strategies for retrieval model training.}  
    We implement a comprehensive data quality improvement framework, combining hard negative mining on structured data with refinement of both positive and negative samples using open-access sources. These strategies substantially improve the effectiveness and robustness of the resulting retrieval models.

    \item \textbf{GeoRAG-QA: a new benchmark for geoscience information retrieval.}  
    We create and release \textit{GeoRAG-QA}, the first large-scale benchmark designed for evaluating geoscience retrieval systems. It contains OA paper-based domain-specific documents and 938 expert-reviewed query-document pairs, providing a standardized framework for measuring performance and guiding future research in this area.

\end{itemize}

\section{GeoGPT.RAG system UI}

\noindent Integrating Retrieval-Augmented Generation (RAG) into GeoGPT significantly enhances the model’s ability to support domain-specific research in geoscience. By retrieving relevant information from a curated, domain-tailored corpus—the \textit{GeoGPT Library}—the system can generate responses that are both accurate and contextually appropriate. This is particularly advantageous for complex queries that require nuanced understanding of specialized terminology and domain-specific concepts.  

\noindent The \textit{GeoGPT Library} include corpus from open access geoscience papers, which was also used for GeoGPT training,  the publishers and journals of the papers are listed here: \href{https://huggingface.co/datasets/GeoGPT-Research-Project/GeoGPT_Training_Data_from_Open-Access_Papers}{GeoGPT Training Data from Open Access Papers}. To access GeoGPT, please visit our website for free \href{https://geogpt.zero2x.org/}{geogpt.zero2x.org}. 

\begin{figure}[h!]
\centering
 \includegraphics[width=0.9\textwidth]{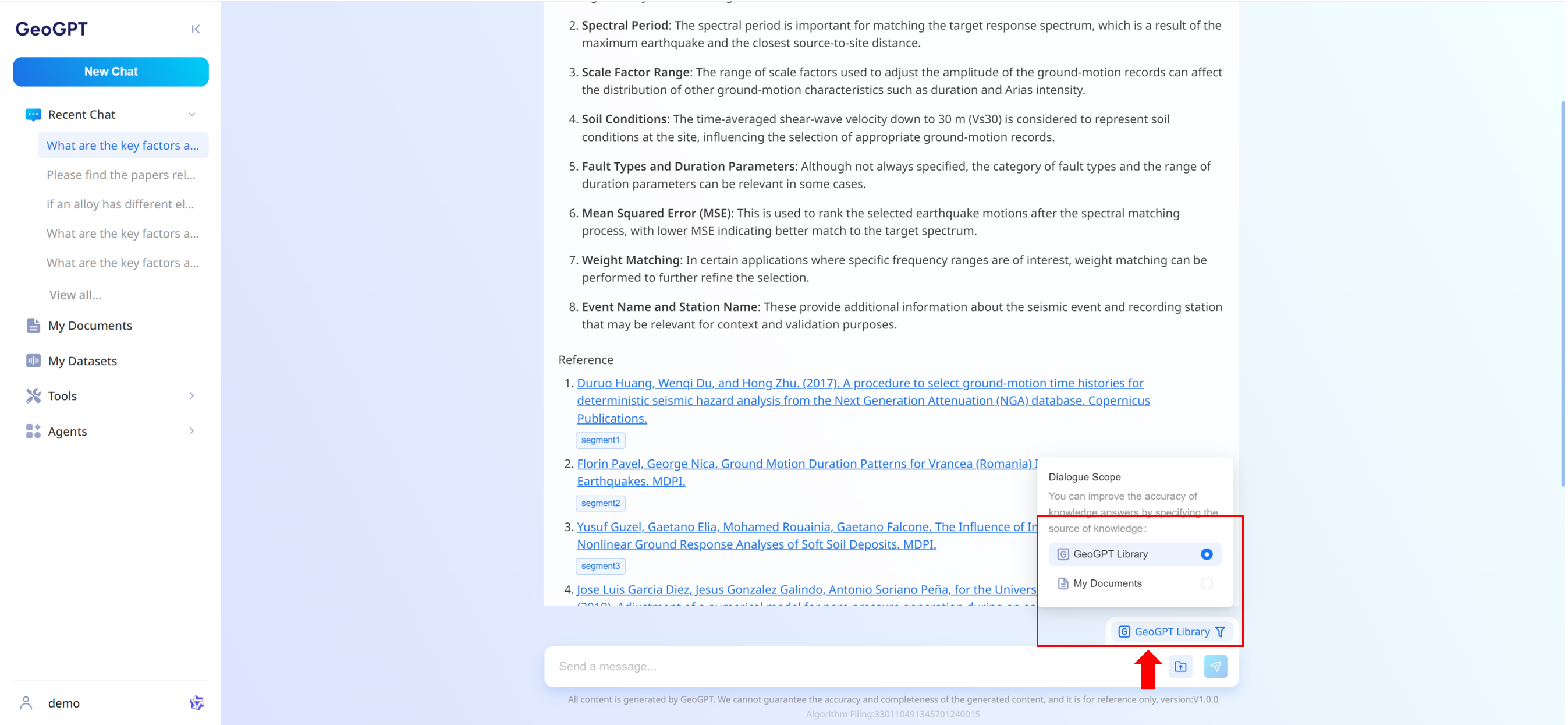}
 \caption{ChatBox Interface for RAG with GeoGPT Library}
 \label{fig6-1-2-0}
\end{figure}

\noindent In addition to the shared library, GeoGPT allows each user to build a personalized knowledge base through the \textit{My Documents} feature. Users can upload their own collections of academic papers or technical documents, which are then integrated into the RAG pipeline for private, customized retrieval. This functionality effectively creates a proprietary version of GeoGPT for each user—fine-tuned to their specific areas of expertise and interests. By leveraging their own data, users can generate content that reflects their unique knowledge and perspectives, improving the relevance, depth, and accuracy of the output. The system also supports continuous updates, allowing users to refine and expand their personal knowledge base over time, ensuring that it remains current and aligned with their evolving research.

\begin{figure}[h!]
\centering
 \includegraphics[width=0.9\textwidth]{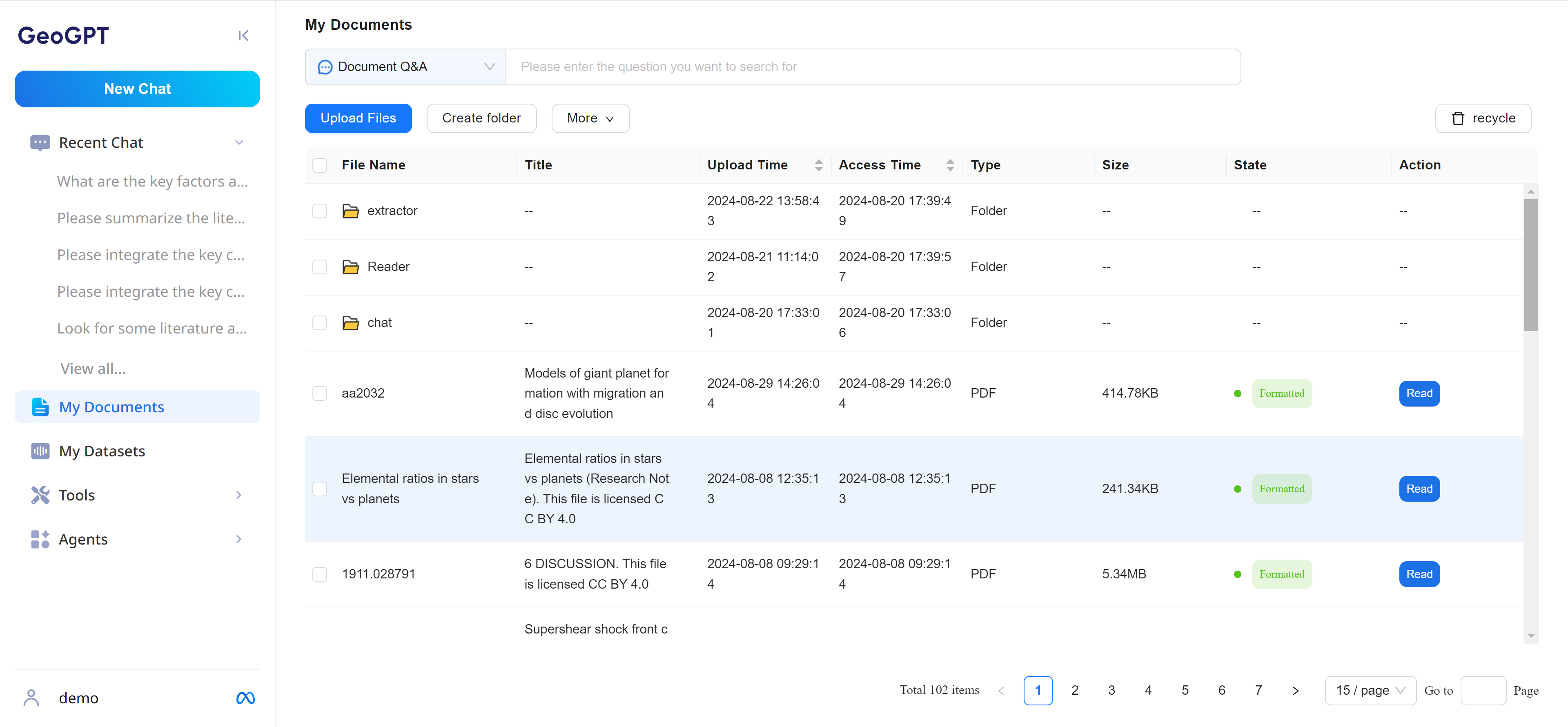}
 \caption{"My Documents" module for documents management}
 \label{fig6-1-2-1}
\end{figure}

\begin{figure}[h!]
\centering
 \includegraphics[width=0.9\textwidth]{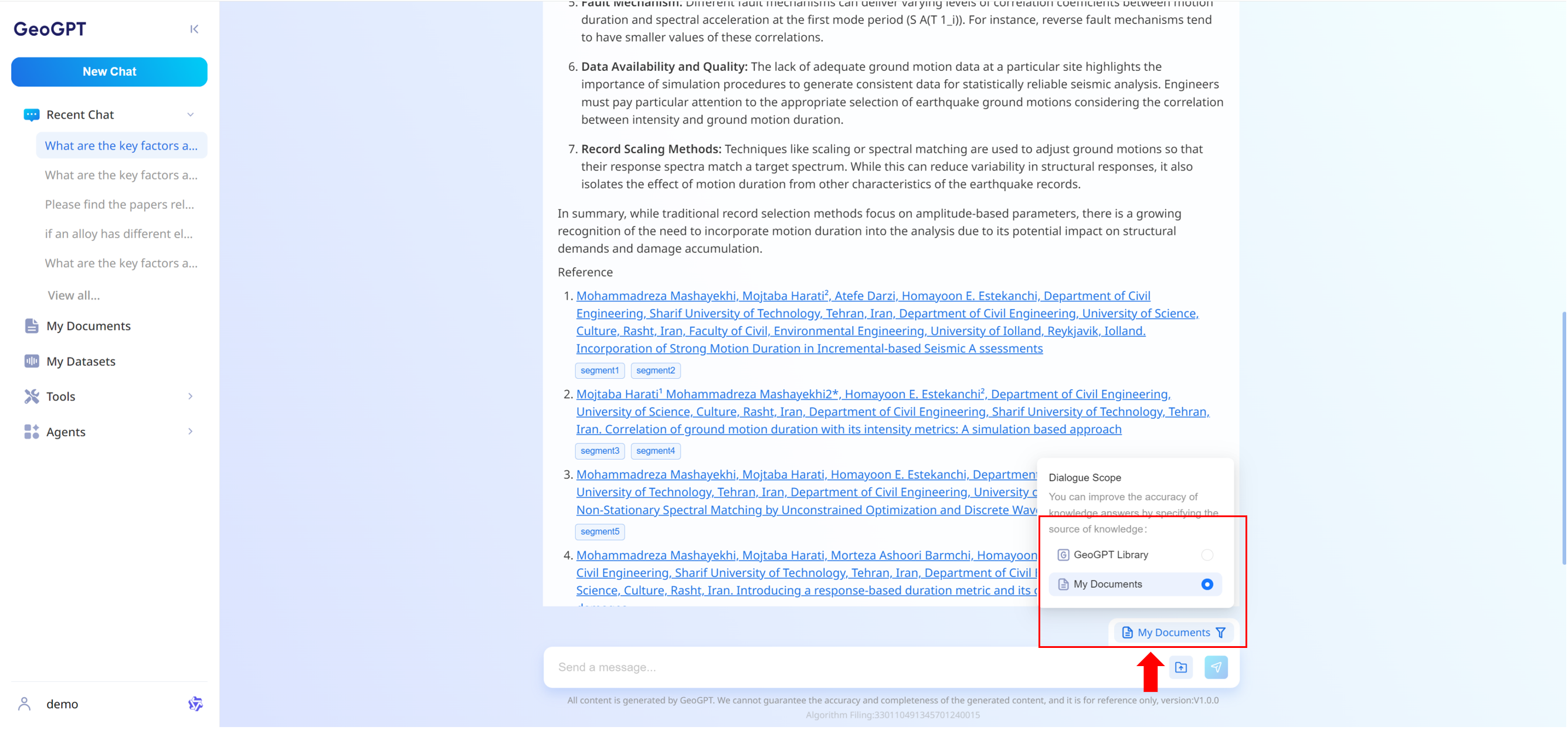}
 \caption{RAG with My Documents}
 \label{fig6-1-2-2}
\end{figure}

\section{RAG Evaluation}
\noindent \textbf{a. Test Set Generation}

\noindent We evaluated GeoGPT’s retrieval and generation capabilities using a custom test set generated with the Test Set Generation module from RAGAS~\cite{es2023ragas}. This test set, named GeoRAG-QA, comprises 1,000 questions derived from publicly available geoscience papers. We used \texttt{gpt-4o} as the underlying LLM for both question-answer (QA) pair generation and answer evaluation, and \texttt{text-embedding-3-small} for calculating embedding-based similarity.

\noindent To manage the computational cost of segment-wise embedding comparisons, we limited the input to a random sample of 1,000 papers from the full corpus of GeoGPT library. During question generation, some entries failed or were incomplete. After manual review, we removed 62 flawed examples, resulting in a final dataset of 938 QA items. These were categorized into four types:

\begin{itemize}
\item 250 single-document, fact-based QA items,
\item 250 single-document inference QA items,
\item 188 multi-document QA items,
\item 250 conditional QA items.
\end{itemize}

\noindent Some generated questions were linked to reference answers marked as “The answer to the given question is not present in the context.” In these cases, we used GPT-4o to regenerate more appropriate reference answers, ensuring evaluation quality and consistency.

\noindent\textbf{b. Retrieval Results}

\noindent Equipped with the knowledge base of GeoGPT library, our retrieval system$(top\_k=8, score\_threshold=0.35)$ achieves:

\begin{table}[h]
    \centering
    \caption{Retrieval Results} 
    \label{table_6.1.3.0}
    \small
        \begin{tabular}{ccc} 
            \toprule
            \textbf{QA Type} & \textbf{Question having retrieved results / All} & \textbf{Avg. Text Score} \\
            \midrule
            Single-document Simple QA & 246/250 & 0.873 \\
            Single-document Inference QA & 246/250 & 0.805 \\
            Multi-document QA & 188/188 & 0.874 \\
            Conditional QA & 247/250 & 0.820 \\
            \midrule
            \textbf{Total} & 927/938 (0.988) & 0.842 \\
            \bottomrule
        \end{tabular}
\end{table}

\begin{table}[h]
    \centering
    \caption {Top recall Results} 
    \label{table_6.1.3.1}
    \small
        \begin{tabular}{ccccccc} 
            \toprule
            \textbf{Retrieval system} & \textbf{Top 1} & \textbf{Top 3} & \textbf{Top 5} & \textbf{Top 8} & \textbf{Top 32} & \textbf{Top 64}\\
            \midrule
            GeoGPT-RAG & 0.908 & 0.945 & 0.95 & 0.959 & 0.966 & 0.969 \\
            \bottomrule
        \end{tabular}
\end{table}

\noindent\textbf{c. End-to-end Evaluation}

\noindent\textbf{Evaluation Metric: Answer Recall} \\
We adopt \textit{Answer Recall} as the primary metric for end-to-end evaluation. This metric quantifies the completeness of generated answers by measuring the proportion of reference (ground-truth) statements that appear in the model-generated response~\cite{es2023ragas}. It reflects the model’s ability to accurately retrieve and synthesize relevant information from the underlying knowledge base.

$$\text{Answer Recall} = \frac{\text{Number of ground-truth statements present in the model's answer}}{\text{Total number of ground-truth statements}}$$

\begin{table}[h]
    \centering
    \caption{Evaluation Results} 
    \label{table_6.1.3.2}
    \small
        \begin{tabular}{ccc} 
            \toprule
            \textbf{Model} & \textbf{Answer Recall} \\
            \midrule
            GeoGPT-0630 & 0.529 \\
            GeoGPT-0630 + RAG & 0.666 \\
            \bottomrule
        \end{tabular}
\end{table}

\noindent The integration of RAG significantly boosts answer completeness. Compared to the base model, GeoGPT enhanced with RAG achieves a 13.7\% absolute gain in Answer Recall, highlighting the effectiveness of retrieval-based augmentation in improving factual grounding.

\vspace{1em}
\noindent\textbf{Expert Evaluation on Real-World Queries} \\
\noindent We further conducted manual evaluation on 70 domain-specific questions related to petroleum exploration. These questions were reviewed by subject matter experts to assess the factual accuracy of the generated answers. The evaluation confirms the robustness of RAG-augmented responses in real-world scenarios.

\begin{table}[h]
    \centering
    \caption{RAG Accuracy} 
    \label{table_6.1.3.3}
    \small
        \begin{tabular}{cccc} 
            \toprule
            \textbf{Model} & \textbf{test data} & \textbf{Evaluation method}& \textbf{answer Accuracy} \\
            \midrule
            GeoGPT+RAG & 70  & Expert & 0.857 \\
            \bottomrule
        \end{tabular}
\end{table}

\noindent Table \ref{table_6.1.3.3} shows the model achieved 85.7\% accuracy on expert-reviewed questions, confirming the practical effectiveness of RAG-enhanced GeoGPT in specialized geoscience domains.

\noindent\textbf{d. Retrieval Augmented Fine Tuning (RAFT) Evaluation}

\noindent To further improve the model’s ability to leverage retrieved content, we incorporated task-specific RAG examples into the supervised fine-tuning (SFT) stage. This strategy—termed \textbf{Retrieval-Augmented Fine-Tuning (RAFT)}—provides the LLM with explicit examples of how to integrate retrieved content during response generation.

\begin{table}[h]
    \centering
    \caption{RAG QA Recall} 
    \label{table_6.1.3.4}
    \small
        \begin{tabular}{ccc} 
            \toprule
            \textbf{Model} & \textbf{In-Domain Recall} & \textbf{Out-Domain Recall} \\
            \midrule
            GeoGPT without RAG-training data & 69.72 & 42.04 \\
            GeoGPT with RAG-training data & 78.12 & 46.49 \\
            \midrule
            \textbf{Advance} & 12.05\% & 10.59\% \\
            \bottomrule
        \end{tabular}
\end{table}

\noindent The inclusion of RAG-style data during the supervised fine-tuning phase—referred to as \textbf{Retrieval-Augmented Fine-Tuning (RAFT)}—led to substantial performance gains. During RAFT, we introduced task-specific examples that demonstrated how retrieved content should be integrated into responses. These examples taught the model to align its generation more effectively with supporting evidence, reinforcing retrieval-aware reasoning during training.

\noindent We evaluated the impact of RAFT on both in-domain (geoscience) and out-of-domain (general) question-answering tasks. As shown in Table~\ref{table_6.1.3.4}, the model trained with RAFT achieved an in-domain recall of 78.12\%, compared to 69.72\% without RAG-style training—a relative improvement of 12.05\%. For out-of-domain tasks, recall increased from 42.04\% to 46.49\%, a 10.59\% gain. These improvements suggest that RAFT enhances the model’s ability to integrate retrieved information into its outputs, leading to more complete, grounded, and contextually accurate responses.

\noindent Overall, the RAFT approach proves highly effective for reinforcing retrieval-augmented behavior during training. It not only boosts domain-specific performance but also improves generalization, making GeoGPT more reliable across both targeted and open-ended information retrieval tasks.

\section{RAG Optimization \& Model Fine-tuning}

\begin{figure}[h!]
\centering
 \includegraphics[width=0.9\textwidth]{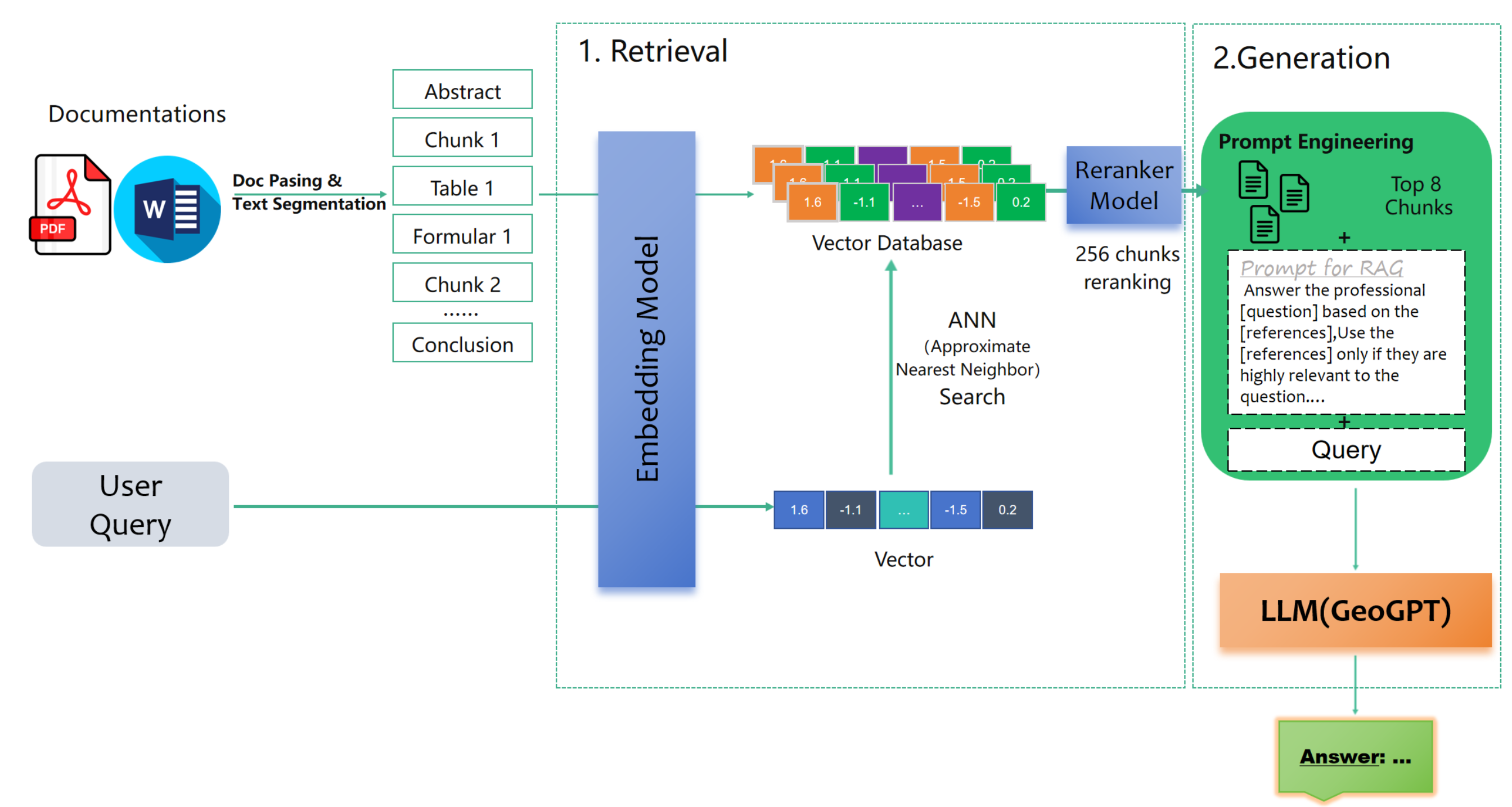}
 \caption{RAG Processing Architecture}
 \label{fig6-1-4-3}
\end{figure}

\noindent To maximize the effectiveness of Retrieval-Augmented Generation (RAG) in GeoGPT, we implemented a multi-stage optimization process. The architecture, illustrated in Figure~\ref{fig6-1-4-3}, supports both batch processing of uploaded PDF files and real-time online retrieval. Our optimization strategy focuses on three core components: text segmentation, embedding model fine-tuning, and reranker model fine-tuning.

\noindent\textbf{a.Text Segmentation}

\noindent Information chunking plays a critical role in enabling effective retrieval within RAG systems. By dividing lengthy documents into semantically coherent segments, we ensure that each chunk can be independently indexed, retrieved, and interpreted by the language model. This step is particularly important for scientific documents, which often contain dense technical language and long paragraphs with multiple concepts.

\noindent To implement chunking, we developed a semantic segmentation pipeline tailored for long-form scientific texts. For input passages exceeding 512 tokens—the typical limit for embedding models—we apply a multi-stage process:
\begin{itemize}
    \item First, the input text is tokenized into individual sentences using NLTK's sentence segmentation toolkit. This creates a candidate list of sentence boundaries.
    \item Next, we apply a BERT-based Next Sentence Prediction (NSP) model to all adjacent sentence pairs. The NSP model outputs a probability score (logit) indicating the likelihood that the second sentence naturally follows the first.
    \item We then normalize these logits and identify the optimal split points by selecting the boundaries with the highest confidence where the predicted NSP label is 1 (i.e., coherent sentence continuation).
    \item Once optimal split points are selected, we segment the text accordingly. If any resulting chunk still exceeds the 512-token threshold, we recursively apply the same NSP-based segmentation strategy until all chunks are within the token limit.
\end{itemize}

\begin{figure}[h!]
\centering
 \includegraphics[width=0.9\textwidth]{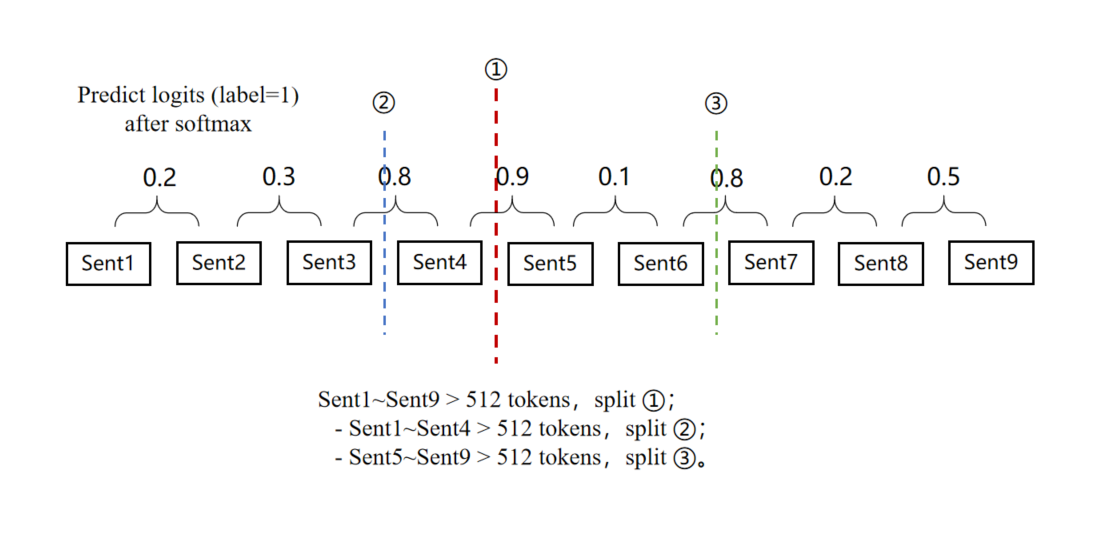}
 \caption{Segmentation with 512 tokens}
 \label{fig6-1-4-0}
\end{figure}

\noindent This method preserves semantic continuity while adhering to token constraints, avoiding arbitrary cuts that could distort meaning. Figure~\ref{fig6-1-4-0} illustrates the final output of this segmentation strategy applied to long paragraphs, showing the alignment between sentence boundaries and token count thresholds.

\noindent\textbf{b. Training Data Generation \& Enhance RAG capability for Large Language Model}

\noindent i) SFT Data Synthesis in Geo Science

\begin{figure}[h!]
\centering
 \includegraphics[width=0.9\textwidth]{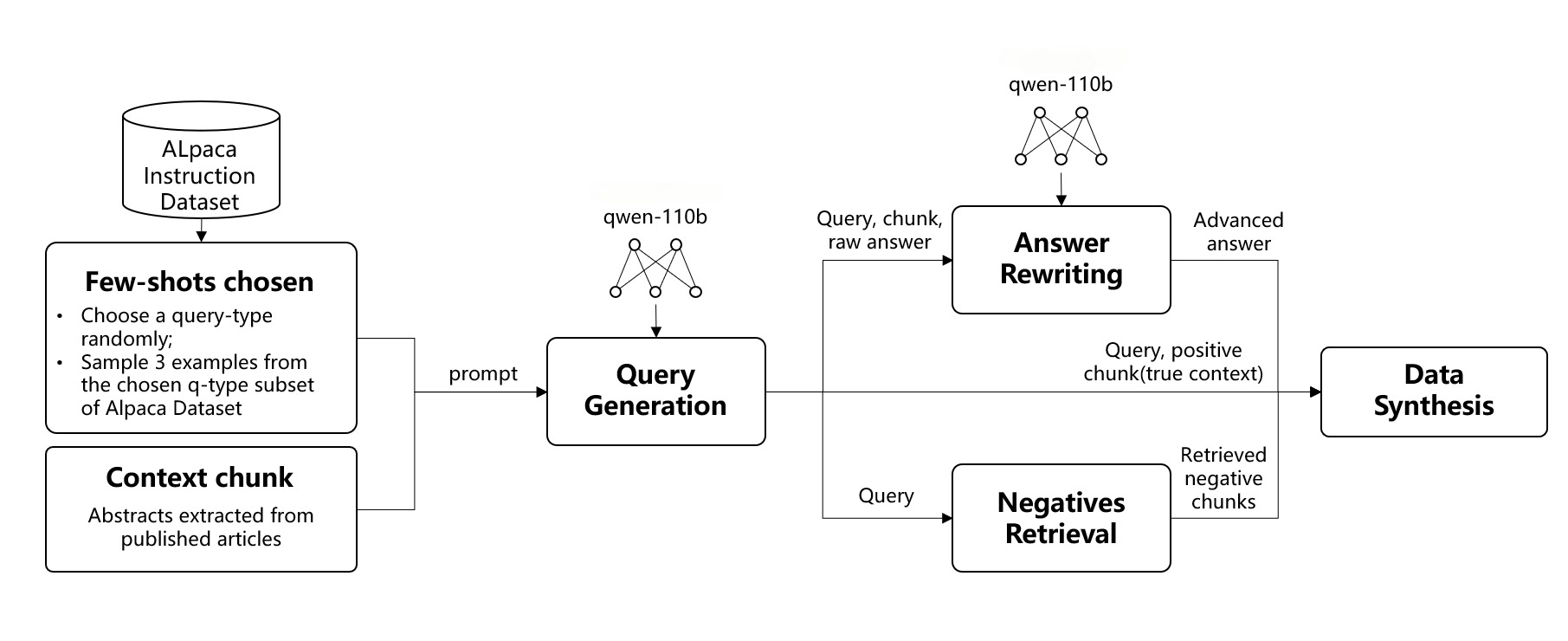}
 \caption{Data Synthesis Process}
 \label{fig6-1-4-1}
\end{figure}

\noindent To generate a diverse set of academic questions in Geoscience, we selected abstracts from one million publicly available papers as the context. We used the Alpaca instruction set from \cite{mecklenburg2024injecting} as a few-shot sampling library, categorizing the instruction set by question types. When constructing the prompt, we first selected a question type randomly and then sampled three instruction data points from the corresponding subset to inject as few-shots into the prompt, guiding the generation of questions for that specific type. This approach increases the variety of question types in the data.
 
\noindent Once the initial generation was complete, we applied an Answer Rewriting Prompt to enhance the completeness and clarity of the answers. Additionally, for each question, we retrieved relevant text chunks from a knowledge base built on all publicly available papers, incorporating these as "negatives" to introduce a certain level of noise. As a result, each data context included both the original context chunk (positive) and supplementary chunks from retrieval, thereby strengthening the model's ability to discern relevant information.
 
\noindent To further refine the model’s ability to distinguish textual information, we added a portion of data (approximately 7:1) where only less relevant, negatively correlated chunks were selected as the context in the retrieval process. In these cases, the answer was reformulated to "Sorry. I cannot find the answer based on the context." as mentioned in \cite{liu2024chatqa}.

\paragraph{Query-Generation Prompt:}
Instruction: Given the next [document], create a [question] and [answer] pair that are grounded in the main point of the document, don't add any additional information that is not in the document and [use prompt by different query type of Table \ref{tab:q_type_prompt}]. The [question] is by an information-seeking user and the [answer] is provided by a helping AI Agent.
Refer to the following question format and corresponding answers. Your output should consist solely of question-answer pairs.

[question]\{few-shot 1 question\}

[answer] \{few-shot 1 answer\}

……

[document]: {contexts}

\#\#\# Response:

\begin{table}[h]
    \centering
    \caption{Query type prompt categories}
    \label{tab:q_type_prompt}
    \small
    \resizebox{\textwidth}{!}{
        \begin{tabular}{ccccccccc}
            \toprule
            \textbf{What} & \textbf{Which} & \textbf{Who/Whose} & \textbf{When} & \textbf{Where} & \textbf{How} & \textbf{Why} & \textbf{General question} & \textbf{Imperative question} \\
            \midrule
            \multicolumn{7}{c}{The question should use \texttt{[{q\_word}]\ldots} to ask} &
            Please ask in general form. &
            Use imperative sentences to prompt the text. \\
            \bottomrule
        \end{tabular}
    }
\end{table}


\paragraph{Answer-Rewriting Prompt:}
Here is a task involving RAG (Retrieval Augmented Generation) for question answering. I will provide you some documents(denoted as [References]), a question (denoted as [Query]) related to the documents, and the corresponding original answer (denoted as [Short Answer]). You are required to expand the content of the answer, with the following requirements:
1. Your generated answer should contain 6 to 8 sentences.
2. Your generated answer should have exactly the same meaning as the [Short Answer] and must perfectly address the [Query] without deviating.
3. The content of your generated answer should fully utilize the content from the [References], and you must not fabricate any facts.
 
[References]
\{documents\_here\}
 
[Query]
\{question\_here\}
 
[Short Answer]
\{short\_answer\_here\}
 
Please note that do not output content other than the generated new answer. Your generated new answer is

\noindent ii)  The Dataset of DPO Processing

\noindent In this study, we used a large language model (LLM) to extract salient information from each question and its corresponding answer, identifying key concepts to guide relevance assessment. We then computed a semantic relevance score between each question and the candidate text chunks retrieved by the Retrieval-Augmented Generation (RAG) system. This scoring was based on the degree of alignment between the extracted key points and the retrieved content. To ensure high-quality training data, we filtered out low-relevance pairs and retained only those question–chunk pairs whose relevance scores exceeded a predefined threshold of 0.4. This threshold was empirically chosen to balance precision and recall in identifying contextually appropriate retrievals for use in subsequent model training.


\begin{figure}[h!]
\centering
 \includegraphics[width=0.9\textwidth]{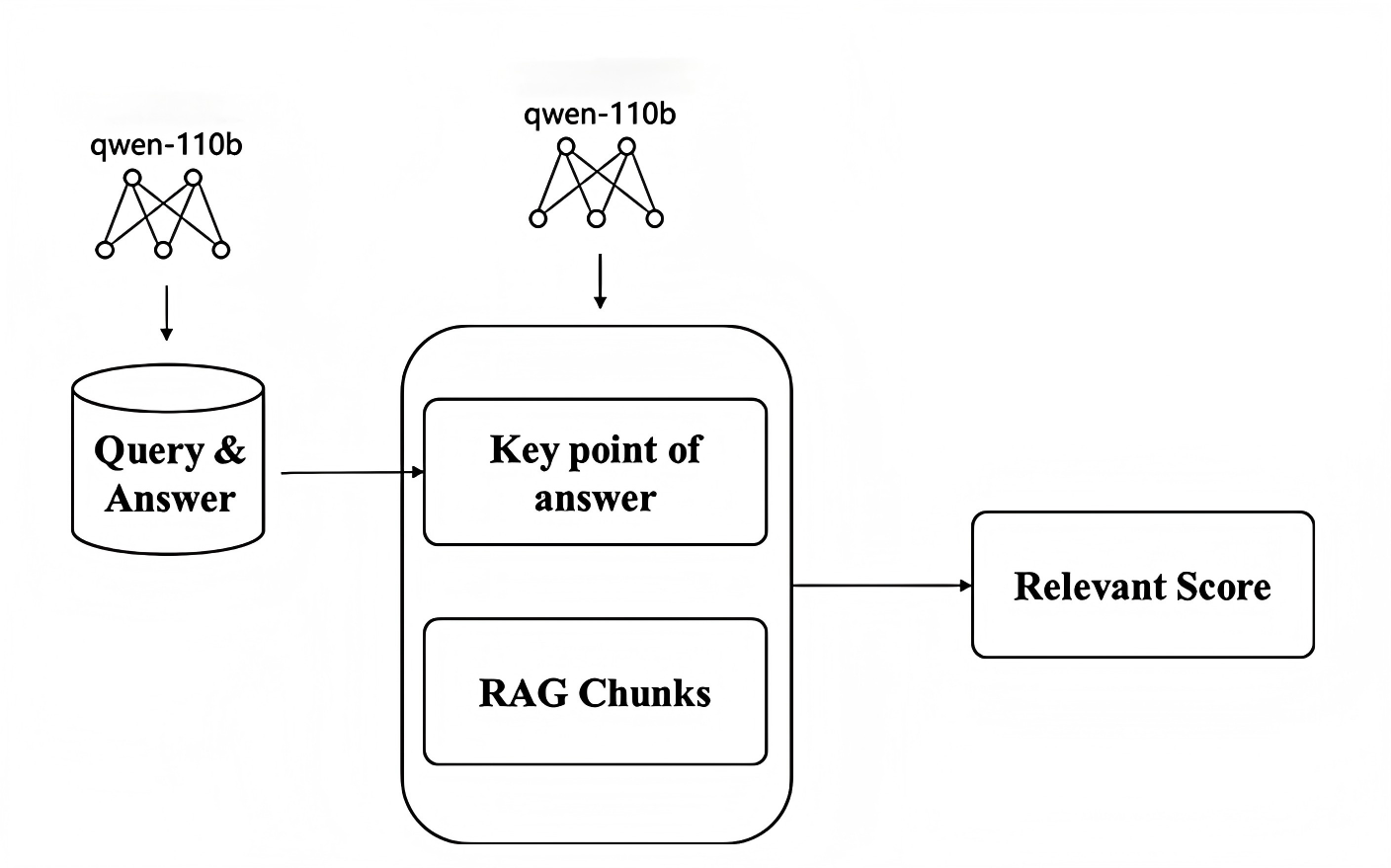}
 \caption{Data Synthesis Process for DPO}
 \label{fig6-1-4-2}
\end{figure}

\noindent\textbf{c. Embedding Model Fine-tuning}

\noindent Embedding models are fundamental components in modern language systems, transforming text into dense, continuous vector representations. These representations support a wide range of downstream tasks, including retrieval, reranking, classification, clustering, and semantic textual similarity. Within RAG frameworks, embedding models play a central role by enabling the retrieval of semantically relevant documents from large external knowledge bases.

\noindent Historically, encoder-based architectures such as BERT and T5 have dominated the landscape of text embedding models. Recent advances in decoder-only transformer models have demonstrated competitive—and in many cases superior—performance for embedding tasks~\cite{SFRAIResearch2024,li2024makingtextembeddersfewshot}. However, studies~\cite{SFR-embedding-2} show that converting decode-only transformer models to encode-only architectures further improves performance. In parallel, several studies~\cite{wang2023improving,lee2024gecko} have shown that large language models can be leveraged to generate high-quality synthetic training data, enabling effective fine-tuning with fewer training steps.

\noindent In this work, we introduce \textbf{GeoEmbedding}, a domain-adapted embedding model fine-tuned from a state-of-the-art encode-only transformer converted from a decode-only transformer. To improve its in-domain retrieval capabilities, we trained GeoEmbedding using synthetic geoscientific data, specifically tailored to the linguistic and semantic patterns found in academic literature from the geography and earth sciences domains.


\noindent \textbf{Datasets.} To train our GeoEmbedding model, we used a combination of publicly available datasets spanning a range of NLP tasks, including retrieval, classification, clustering, reranking, and semantic textual similarity (STS). This diverse set of tasks enables the model to learn general-purpose embeddings while being robust to domain-specific requirements. We sample about 360k data from datasets including SciFact~\cite{wadden-etal-2020-fact}, S2ORC~\cite{lo-wang-2020-s2orc}, PAQ~\cite{lewis2021paq}, Specter~\cite{cohan2020specter}, and FiQA~\cite{maia201818}, etc. Additionally, we incorporate some datasets from the MTEB benchmark, such as AmazonCounterfactual, Banking77, Emotion, MassiveIntent, MassiveScenario, and ToxicConversations, etc.



\noindent Following standard practice in prior work such as SFR-Embedding-Mistral and NV-Embed, we train using only the training splits from each dataset. Where datasets include annotated negatives, we use them directly for contrastive learning. For datasets without sufficient negative samples—or none at all—we apply SimANS~\cite{zhou2022simans} to mine hard negatives from the corresponding corpus.

\noindent In addition to general-purpose datasets, we generate domain-specific data for geoscience using Qwen. These synthetic examples are based on abstracts from open-access scholarly articles collected from arXiv under CC BY license, with proper attribution. We compile these into a dedicated training set, and apply SimANS to mine hard negatives from the domain corpus, ensuring the model learns to distinguish subtle differences in geoscientific content.

\noindent \textbf{Training.}  Given the diversity of tasks across datasets, we prepend a task-specific instruction to each query to guide the model during training. For each query–document pair, we construct an \textit{instructed query} by combining the task instruction $I$ with the query $q$ in the following format:
$$q_\text{inst} = \text{Instruct: }\{I\} \backslash \text{nQuery: }\{q\}$$
\noindent We apply this formatting only to the query; no instruction is added to the document side. 
To fine-tune the embedding model, we adopt instruction tuning with in-batch negatives using the InfoNCE loss function:
$$\min \mathbb L=-\log \frac{\phi(q_\mathrm{inst},d^+)}{\phi(q_\mathrm{inst},d^+)+\sum_{n_i\in N}\phi(q_\mathrm{inst},n_i)}$$
\noindent where $q_\text{inst}$ is the instructed query, $d^+$ is the corresponding positive document, and $n_i$ denotes the $i$-th negative document in the batch. The similarity function $\phi(\cdot, \cdot)$ is defined using a temperature-scaled cosine similarity:
$$\phi(q,d)=\exp \left(\frac{\cos(\mathbf h_q,\mathbf h_d)}{\tau}\right)$$
\noindent where $\mathbf{h}_q$ and $\mathbf{h}_d$ are the vector embeddings of the query and document, respectively, and $\tau$ is a temperature hyperparameter controlling the sharpness of the similarity distribution.

\noindent To enable parameter-efficient training, we apply Low-Rank Adaptation (LoRA)~\cite{hu2021lora}, setting the rank $r=8$ and scaling factor $\alpha=16$. The model is fine-tuned from \textbf{Mistral-7B-v0.1}~\cite{Jiang2023Mistral7}, a 7B decoder-only transformer pretrained model that achieved state-of-the-art performance in 2023. We fine-tune the model over 4096 gradient accumulation steps with a learning rate of $1 \times 10^{-5}$. 

\noindent We convert the Mistral model into an encode-only architecture by removing its causal attention masks. Additionally, we apply Matryoshka representation learning during training to enhance the quality of low-dimensional embeddings.


\noindent \textbf{Evaluation.} We evaluate our embedding model on both general-purpose and domain-specific benchmarks to assess its performance across retrieval tasks.

\begin{figure}
    \centering
    \includegraphics[width=0.8\textwidth]{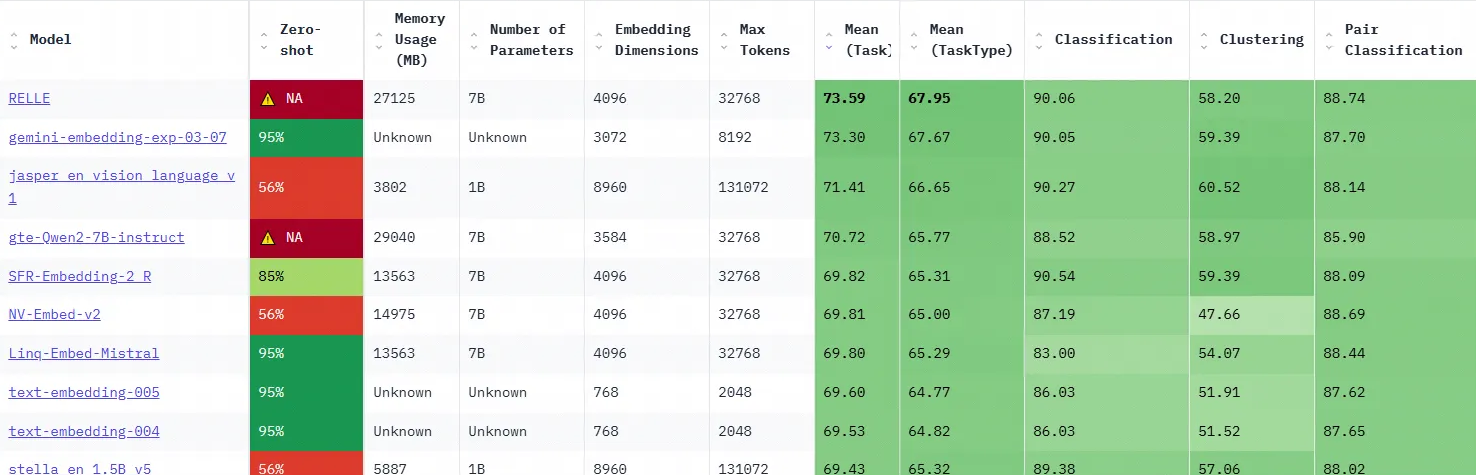}
    \caption{MTEB (eng) benchmark as of April 22, 2025.}
    \label{fig:mteb_eng_1}
\end{figure}

\noindent For general evaluation, we use the \textbf{MTEB(eng)} benchmark, which includes a diverse set of retrieval datasets. Table~\ref{tab:embedding_eval_mteb_all} summarizes our results. We compare our model against two strong baselines—SFR-Embedding-Mistral and NV-Embed-v2—using scores replicated from the official MTEB leaderboard\footnote{\url{https://huggingface.co/spaces/mteb/leaderboard}} (visualized in Figure~\ref{fig:mteb_eng_1}). Despite being fine-tuned on synthetic in-domain data, our model retains competitive performance across general retrieval benchmarks, demonstrating robust generalization ability.

\begin{figure}[h]
    \centering
    \begin{tikzpicture}
        \begin{axis}[
            xlabel={Top-$k$},
            ylabel={Recall},
            legend pos=south east,  
            grid=major,
            xmin=0, xmax=64,
            ymin=0.7, ymax=1.0
        ]
        \addplot coordinates {(1,0.7145) (3,0.9021) (5,0.9375) (10,0.9626) (32,0.9888) (64,0.9916)};
        \addplot coordinates {(1,0.7453) (3,0.9216) (5,0.9514) (10,0.9701) (32,0.9869) (64,0.9916)};
        \legend{SFR-Embedding-Misral, GeoEmbedding}  
        \end{axis}
    \end{tikzpicture}
    \caption{Top-$k$ recall on GeoRAG-QA benchmark}
    \label{fig:embedding_eval_topk_recall}
\end{figure}
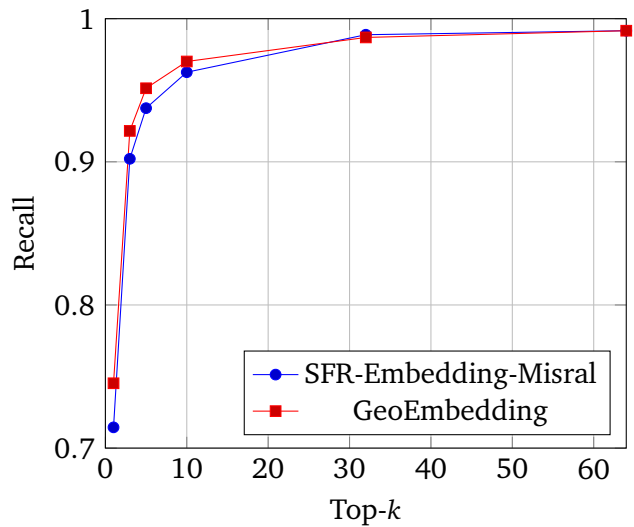

\noindent To evaluate in-domain performance, we constructed a dedicated geoscience retrieval benchmark using \textbf{GeoRAG-QA}. This benchmark includes  synthetic query–document pairs generated using Qwen-110B and a retrieval corpus of GeoGPT library. As shown in Figure~\ref{fig:embedding_eval_topk_recall}, our fine-tuned model achieves substantial gains in retrieval accuracy across all top-$k$ settings—from Recall@1 through Recall@10—demonstrating its effectiveness in domain-specific information retrieval.


\begin{table}
    \centering
    \caption{Evaluation on MTEB tasks}
    \label{tab:embedding_eval_mteb_all}
    \resizebox{\textwidth}{!}{
        \begin{tabular}{ccccccccc}
            \toprule
            & \textbf{Classification} & \textbf{Clustering} & \textbf{PairClassification} & \textbf {Reranking} & \textbf{Retrieval} & \textbf{STS} & \textbf{Summarization} & \textbf{Overall} \\
            \midrule
            \textbf{SFR-Embedding-Mistral} & 80.47 & 54.93 & 88.59 & \textbf{50.15} & 59.33 & 84.77 & 36.32 & 69.31 \\
            \textbf{NV-Embed-V2} & 87.19 & 47.66 & \textbf{88.69} & 49.61 & \textbf{62.84} & \textbf{83.82} & \textbf{35.21} & 69.81 \\
            \textbf{GeoEmbedding} & \textbf{89.67} & \textbf{56.50} & 82.51 & 48.32 & 60.91 & 80.60 & 30.43 & \textbf{70.21} \\
            \bottomrule
        \end{tabular}
    }
\end{table}

\noindent \textbf{d. Reranker Model Fine-tuning}

\noindent The reranker is a crucial component in Retrieval-Augmented Generation (RAG) systems. It refines the initial set of retrieved candidates by reordering them based on their semantic relevance to the query. Unlike embedding models, which encode queries and documents independently into fixed-length vectors, the reranker jointly processes the query and document as input and directly predicts a relevance score. By modeling token-level interactions between the query and document, the reranker captures richer semantic relationships, resulting in improved ranking of the top candidates. Given a query-document pair, the model outputs a raw relevance score, which is typically passed through a sigmoid function to produce a normalized value in the range $[0, 1]$.


\noindent \textbf{Model and Training methods.}  We adopt an industry-standard cross-encoding scheme for reranking and use the BGE-M3 model~\cite{chen-etal-2024-m3} as our base. BGE-M3 is known for its strong generalization capabilities and is supported by a robust open-source ecosystem, making it well-suited for real-time applications.

\noindent During training, we feed both positive and negative samples corresponding to each query into the model simultaneously. To optimize ranking performance, we apply the Multiple Negative Ranking (MNR) loss, defined as:
$$\mathcal{L}_{\text{MNR}} = -\frac{1}{N} \left( \sum_{i=1}^C \log(p_{\text{neg}_i}) - \log(p_{\text{pos}}) \right)$$
where $N$ is the total number, $C$ is the number of negative samples per query, $p_{\text{pos}}$ is the predicted relevance score for the positive document, and $p_{\text{neg}_i}$ is the score for the $i$-th negative document. This loss encourages the model to rank the positive document higher than its negative counterparts, thereby improving retrieval precision in downstream tasks.


\textbf{Dataset and Quality Improvement.} To maximize the reranker's performance in the geoscience domain, we curated a large training corpus composed of both open-source and domain-specific ranking pairs. The domain-specific data were synthesized using Qwen, conditioned on abstracts from open-access scholarly articles in geography sourced from arXiv. 

\noindent After assembling all positive query–document pairs, we applied SimANS~\cite{zhou2022simans} to mine hard negative samples from the candidate corpus, thus forming a balanced and challenging training dataset. Since both synthetic and open-source data were generated in batches, individual data points may vary in quality. Even a small number of noisy or biased examples can significantly degrade reranking performance. To address this, we implemented a data quality filtering process to improve the reliability of training examples. For each positive pair, we used the LLaMA3 model to assess answer quality based on a set of predefined criteria. Each sample was assigned a quality score ranging from 0 to 3, following the standards detailed in Table~\ref{table_6.1.4.0}. Only samples with sufficient quality were retained for training.Data with a quality label of 0 is excluded from training.

\begin{table}[h]
    \centering
    \caption{data quality standards} 
    \label{table_6.1.4.0}
    \tiny
    \resizebox{\textwidth}{!}{
        \begin{tabular}{cc} 
            \toprule
            \textbf{quality label} & \textbf{quality standards} \\
            \midrule
            3 & completely correct, detailed, and covers all key information with clear and easy-to-understand expression. \\
			2 & mostly correct, covers most key information, and is clearly expressed but with minor shortcomings. \\
			1 & generally correct but has some errors or omissions, with mostly clear expression. \\
			0 & significant errors or omissions, lacks sufficient detail, and is not clearly expressed. \\
            \bottomrule
        \end{tabular}
    }
\end{table}

\noindent We ultimately utilized diverse types of paired data to train the model. Open-source data included:

QA pairs: DuReader~\cite{he2017dureader}, NQ, LCQMC~\cite{liu2018lcqmc}, HotpotQA, etc.

Retrieval-oriented pairs: T2Ranking~\cite{xie2023t2ranking}, mMARCO, MS MARCO, etc.

Similarity task pairs: NLI~\cite{koreeda2021contractnli}, NLI-zh, AFQMC~\cite{lee2022twenty}, PAWSX~\cite{yang2019paws}, etc.

Multilingual data: Mr. TyDi.

\noindent In total, we incorporated over 360,000 open-source data pairs, along with 30,000 synthetically generated QA pairs specifically tailored for geological science applications.

\noindent For training with contrastive objectives, negative sample quality plays a critical role in reranker performance. Whenever available, we retain the original negative samples provided within each dataset. However, many datasets either lack negatives entirely or contain an insufficient number of challenging examples. To ensure training robustness, we construct a set of at least 16 negative samples per query for all datasets.

\noindent When additional negatives are required, we use a weak retriever to select hard negative candidates from a large-scale public Wiki corpus or domain-specific geoscience libraries curated from scholarly literature. We then apply the SimANS sampling strategy~\cite{zhou2022simans} to prioritize harder examples from the candidate pool. Each negative sample $neg_x$ is assigned a sampling probability based on the squared difference between its retriever score and that of the corresponding positive sample:
$$\mathcal{F}_{\text{Sample}}(neg_{x}) = \exp\left( -\sigma \cdot (neg_{\text{score}} - pos_{\text{score}})^2 \right)$$
\noindent where $neg_{\text{score}}$ and $pos_{\text{score}}$ are the scores returned by the weak retriever, and $\sigma$ is a task-specific hyperparameter that adjusts the sharpness of the sampling distribution. $\mathcal{F}_{\text{Sample}}(neg_{x})$ represents the probability of selecting $neg_x$ as a training negative.

\noindent We adopt different $\sigma$ values to reflect the difficulty characteristics of various dataset types. For question-answering datasets—where semantic precision is critical—we use a sharper distribution with $\sigma = 1$, focusing the model on distinguishing subtle contextual differences. For reranking tasks and semantic similarity datasets, where broader generalization is beneficial, we use a more relaxed sampling distribution with $\sigma = 3$.

\noindent By ensuring a consistent quantity of informative and diverse negative samples, this strategy strengthens the model's ability to discriminate between highly similar documents and improves ranking accuracy in both general and domain-specific settings.

\noindent \textbf{Evaluation.} We adopted a dual evaluation approach assessing both domain-specific and general capabilities. For domain-specific performance, we conducted top-k recall tests using domain-specific datasets, with the results presented in Table \ref{table_6.1.3.1}. For general capabilities, we evaluated the model on the BEIR\cite{thakur2021beir} benchmark dataset, which include includes 16 rank evaluation datasets from different domains. Under the condition that all retrievers used BM25, our model outperformed open-source state-of-the-art model in general retrieval tasks as well.

\begin{table}[h]
    \centering
    \caption{BEIR General Benchmark metric} 
    \label{table_6.1.4.2}
    \small
        \begin{tabular*}{0.7\textwidth}{@{\extracolsep{\fill}}ccc}  
            \toprule
            \textbf{BEIR metric} & \textbf{bge-reranker-v2-m3} & \textbf{GeoReranker} \\
            \midrule
                 \textbf{Map@3 average}  & 0.31295 & 0.317110667 \\
            \textbf{NDCG@10 average} & 0.468298125 & 0.477822667 \\
            \bottomrule
        \end{tabular*}
\end{table}

\section{RAG Engineering}

\noindent \textbf{a. Vector Database}

\noindent The vector database used in this RAG system is deployed via Zilliz Cloud, a fully managed service provided by Zilliz, the original creators of Milvus. Zilliz Cloud offers a deeply optimized, out-of-the-box Milvus experience, combining high performance with improved cost-effectiveness and user-friendly deployment. Specifically designed for production-level use, Zilliz Cloud enables application deployment across multiple public cloud environments worldwide, while ensuring enterprise-grade data security and system elasticity. Compared to the open-source Milvus, Zilliz Cloud provides superior support for cloud-native features, scalability, and operational simplicity. A detailed comparison between the two is provided in Table~\ref{table_6.1.7.0}.


\begin{table}[h]
    \centering
    \caption{Vector Database: Milvus VS Zilli Cloud} 
    \label{table_6.1.7.0}
    \small
    \resizebox{\textwidth}{!}{
        \begin{tabular}{ccc} 
            \toprule
             & Milvus & Zilliz Cloud \\
            \midrule
            \textbf{Fully Managed}\\ Milvus clusters that are fully managed and automated with zero operations required. & NO & YES \\
            \textbf{High Availability}\\ 99.95\% uptime SLA with built-in failover to ensure your Milvus clusters are always available. & NO & YES \\
            \textbf{Elastic Scaling}\\ Easily scale up to a billion-scale or down without the need to over-provision infrastructure. & NO & YES \\
            \textbf{Different Machine Types for Best Price Performance}\\ Provides two compute unit (CU) types: performance-optimized CU and capacity-optimized CU to fit different use cases. & NO & YES \\
            \textbf{Infinite Storage}\\ Cost-effectively store data at any scale without the need to increase compute resources. & NO & YES \\
            \textbf{Cloud UI}\\User-friendly GUI to easily manage and monitor your Milvus clusters at any scale in the cloud. & NO & YES \\
            \textbf{Resource monitors}\\Get automatic notifications to avoid service overloading. & NO & YES \\
            \bottomrule
        \end{tabular}
    }
\end{table}

\noindent The RAG system maintains two types of vector databases: a \textit{public library} and a \textit{user library}. The public library is designed for a broad community of scientific researchers and provides professional responses based on academic literature and Wikipedia-derived content. The user library, by contrast, supports personalized document-based question answering. In this setting, users upload their own documents, which the system automatically shards, vectorizes, and stores. This enables users to query their own content in a retrieval-augmented manner.

\noindent To support efficient querying in the user library, the system utilizes the \textit{Partition} feature of Zilliz Cloud. A Partition is a logical component derived from a Collection and allows the physical storage to be divided into multiple segments, each referred to as a Partition. By scoping queries to a specific Partition rather than scanning the entire Collection, the system significantly improves query efficiency. In our implementation, each user's data is stored in its own Partition, using the user ID as the Partition key.

\noindent Currently, the public library contains approximately 15 million vector entries, while the user library comprises around 3 million entries. Zilliz Cloud provides millisecond-level query latency and ultra-high recall rates, ensuring that the RAG system can deliver fast and accurate retrieval for both general-purpose and user-specific queries.


\noindent \textbf{b. Batch Processing of PDF Files from Papers and Books}

\noindent Constructing a comprehensive and high-quality knowledge base is essential for enabling accurate and efficient retrieval in large language model systems based on RAG. PDF files—particularly academic papers, technical books, and scanned documents—are among the most valuable data sources due to their rich textual content and embedded metadata. This section outlines the technical workflow for batch processing PDF files, encompassing file parsing, data extraction, and vectorization for semantic indexing.

\vspace{0.5em}
\noindent\textit{i) File Parsing}

\noindent \textbf{Text Extraction:} Optical Character Recognition (OCR) is applied to recognize and extract text from image-based or scanned PDFs. For digitally generated PDFs, native extraction tools such as PDFPlumber or PyMuPDF are preferred to maintain structural fidelity and avoid OCR errors.
    
\noindent \textbf{Layout Analysis:} To preserve document semantics, the layout of each page is analyzed to identify the spatial organization of elements such as body text, images, captions, tables, equations, and footnotes. Tools like LayoutParser can assist in detecting layout regions and determining content flow across multi-column formats.

\noindent \textbf{Metadata Extraction:} Bibliographic metadata—such as author names, paper titles, publication dates, DOIs, and journal information—is extracted using a combination of embedded XMP metadata and heuristic parsing of document headers. These metadata fields are used to support indexing, filtering, and citation tracking within the knowledge base.

\vspace{0.5em}
\noindent\textit{ii) Data Extraction}

\noindent After parsing, the system performs structured data extraction and cleaning:

\noindent \textbf{Text Cleaning:} The raw extracted text often contains noise such as page headers, footers, watermarks, or extraneous whitespace. These artifacts are removed using rule-based filters and page-aware segmentation. Empty or near-empty pages are discarded.

\noindent \textbf{Structured Data Extraction:} Tabular data is detected and converted into machine-readable formats such as CSV. Paragraphs are segmented and mapped to their logical context (e.g., sections or chapters) based on layout cues and lexical markers. This facilitates document reconstruction and targeted retrieval.

\noindent \textbf{Key Information Identification:} Important content segments such as abstracts, introductions, conclusions, author affiliations, and keywords are automatically identified using pattern matching and model-based classifiers. These sections can later be prioritized during ranking or retrieval.

\vspace{0.5em}
\noindent\textit{iii) Vectorization}

\noindent In the final stage, the cleaned and structured content is prepared for integration into the vector database:

\noindent \textbf{Data Formatting:} Extracted content and associated metadata are serialized into a standardized format, such as JSON or structured key-value pairs, which aligns with the schema expected by the downstream indexing system.

\noindent \textbf{Text Chunking:} To maintain semantic coherence while accommodating input length limitations of embedding models, long text blocks are segmented into overlapping or context-aware chunks. Chunking strategies are optimized to preserve sentence boundaries and thematic continuity.

\noindent \textbf{Data Ingestion:} Each chunk is passed through a pretrained embedding model to generate dense vector representations. These vectors are indexed in the vector database (e.g., Zilliz Cloud or Milvus), along with their corresponding metadata. This enables fast, semantic search during inference and supports scalable document-level and passage-level retrieval.

\noindent \textbf{c. Online Chat}

\noindent To enable interactive question answering grounded in a knowledge base, the system adopts a RAG framework. This approach enhances the performance of LLMs by incorporating relevant external information into the generation process. The online chat pipeline consists of the following steps:

\vspace{0.5em}
\noindent\textit{i) Query Vectorization}

\noindent When a user submits a query, the system first encodes it using a pre-trained embedding model. This transforms the natural language input into a high-dimensional vector representation that resides in the same semantic space as the knowledge base entries.

\vspace{0.5em}
\noindent\textit{ii) Retrieval from the Vector Database}

\noindent The system searches the vector database using the query vector to identify the top-$N$ most similar text chunks. This retrieval step is based on vector similarity (e.g., cosine similarity or inner product) between the query and document embeddings.

\vspace{0.5em}
\noindent\textit{iii) Scoring and Ranking}

\noindent The $N$ retrieved chunks are passed to a ranking model that assigns relevance scores based on their semantic alignment with the user query. The top-$K$ highest-scoring chunks are selected as final retrieval results. This ranking process helps surface the most contextually relevant passages.

\vspace{0.5em}
\noindent\textit{iv) Prompt Construction}

\noindent A prompt is constructed by combining the original user query with the top-$K$ retrieved chunks. The prompt follows a structured format designed to guide the language model in generating accurate and grounded responses. Domain-specific instructions may also be included.

\vspace{0.5em}
\noindent\textit{v) Answer Generation}

\noindent The constructed prompt is fed into a domain-adapted large language model, which synthesizes a response based on both the user query and the retrieved context. The final output aims to be fluent, factually grounded, and semantically relevant to the user’s question.

\section{Conclusion}
\noindent In this technical report, we introduced a RAG-based system designed to advance geoscientific research within the GeoGPT framework. The system integrates a complete data processing workflow, robust RAG engineering, a domain-specific embedding model, a domain-specific reranker model, a data synthesis approach, and strategies for enhancing data quality. We conducted comprehensive, multi-level evaluations, all of which demonstrated significant performance improvements. To foster community development and accelerate geoscientific innovation, we have open-sourced the system code, GeoRAG-QA dataset, GeoEmbedding model, and GeoReranker model.

\newpage
\bibliography{ref}

\begin{thebibliography}{26}
\providecommand{\natexlab}[1]{#1}
\providecommand{\url}[1]{\texttt{#1}}
\expandafter\ifx\csname urlstyle\endcsname\relax
  \providecommand{\doi}[1]{doi: #1}\else
  \providecommand{\doi}{doi: \begingroup \urlstyle{rm}\Url}\fi

\bibitem[Chen et~al.(2024)Chen, Xiao, Zhang, Luo, Lian, and Liu]{chen-etal-2024-m3}
Jianlyu Chen, Shitao Xiao, Peitian Zhang, Kun Luo, Defu Lian, and Zheng Liu.
\newblock {M}3-embedding: Multi-linguality, multi-functionality, multi-granularity text embeddings through self-knowledge distillation.
\newblock In Lun-Wei Ku, Andre Martins, and Vivek Srikumar, editors, \emph{Findings of the Association for Computational Linguistics: ACL 2024}, pages 2318--2335, Bangkok, Thailand, August 2024. Association for Computational Linguistics.
\newblock \doi{10.18653/v1/2024.findings-acl.137}.
\newblock URL \url{https://aclanthology.org/2024.findings-acl.137/}.

\bibitem[Cohan et~al.(2020)Cohan, Feldman, Beltagy, Downey, and Weld]{cohan2020specter}
Arman Cohan, Sergey Feldman, Iz~Beltagy, Doug Downey, and Daniel~S Weld.
\newblock Specter: Document-level representation learning using citation-informed transformers.
\newblock \emph{arXiv preprint arXiv:2004.07180}, 2020.

\bibitem[Es et~al.(2023)Es, James, Espinosa-Anke, and Schockaert]{es2023ragas}
Shahul Es, Jithin James, Luis Espinosa-Anke, and Steven Schockaert.
\newblock Ragas: Automated evaluation of retrieval augmented generation.
\newblock \emph{arXiv preprint arXiv:2309.15217}, 2023.

\bibitem[Gao et~al.(2023)Gao, Xiong, Gao, Jia, Pan, Bi, Dai, Sun, Wang, and Wang]{gao2023retrieval}
Yunfan Gao, Yun Xiong, Xinyu Gao, Kangxiang Jia, Jinliu Pan, Yuxi Bi, Yi~Dai, Jiawei Sun, Meng Wang, and Haofen Wang.
\newblock Retrieval-augmented generation for large language models: A survey.
\newblock \emph{arXiv preprint arXiv:2312.10997}, 2023.

\bibitem[He et~al.(2017)He, Liu, Liu, Lyu, Zhao, Xiao, Liu, Wang, Wu, She, et~al.]{he2017dureader}
Wei He, Kai Liu, Jing Liu, Yajuan Lyu, Shiqi Zhao, Xinyan Xiao, Yuan Liu, Yizhong Wang, Hua Wu, Qiaoqiao She, et~al.
\newblock Dureader: a chinese machine reading comprehension dataset from real-world applications.
\newblock \emph{arXiv preprint arXiv:1711.05073}, 2017.

\bibitem[Hu et~al.(2021)Hu, Shen, Wallis, Allen-Zhu, Li, Wang, Wang, and Chen]{hu2021lora}
Edward~J Hu, Yelong Shen, Phillip Wallis, Zeyuan Allen-Zhu, Yuanzhi Li, Shean Wang, Lu~Wang, and Weizhu Chen.
\newblock Lora: Low-rank adaptation of large language models.
\newblock \emph{arXiv preprint arXiv:2106.09685}, 2021.

\bibitem[Jiang et~al.(2023)Jiang, Sablayrolles, Mensch, Bamford, Chaplot, de~Las~Casas, Bressand, Lengyel, Lample, Saulnier, Lavaud, Lachaux, Stock, Scao, Lavril, Wang, Lacroix, and Sayed]{Jiang2023Mistral7}
Albert~Qiaochu Jiang, Alexandre Sablayrolles, Arthur Mensch, Chris Bamford, Devendra~Singh Chaplot, Diego de~Las~Casas, Florian Bressand, Gianna Lengyel, Guillaume Lample, Lucile Saulnier, L'elio~Renard Lavaud, Marie-Anne Lachaux, Pierre Stock, Teven~Le Scao, Thibaut Lavril, Thomas Wang, Timoth{\'e}e Lacroix, and William~El Sayed.
\newblock Mistral 7b.
\newblock \emph{ArXiv}, abs/2310.06825, 2023.
\newblock URL \url{https://api.semanticscholar.org/CorpusID:263830494}.

\bibitem[Koreeda and Manning(2021)]{koreeda2021contractnli}
Yuta Koreeda and Christopher~D Manning.
\newblock Contractnli: A dataset for document-level natural language inference for contracts.
\newblock \emph{arXiv preprint arXiv:2110.01799}, 2021.

\bibitem[Lee et~al.(2024)Lee, Dai, Ren, Chen, Cer, Cole, Hui, Boratko, Kapadia, Ding, et~al.]{lee2024gecko}
Jinhyuk Lee, Zhuyun Dai, Xiaoqi Ren, Blair Chen, Daniel Cer, Jeremy~R Cole, Kai Hui, Michael Boratko, Rajvi Kapadia, Wen Ding, et~al.
\newblock Gecko: Versatile text embeddings distilled from large language models.
\newblock \emph{arXiv preprint arXiv:2403.20327}, 2024.

\bibitem[Lee et~al.(2022)Lee, Pham, and Reichman]{lee2022twenty}
Joonho Lee, Hung~Q Pham, and David~R Reichman.
\newblock Twenty years of auxiliary-field quantum monte carlo in quantum chemistry: An overview and assessment on main group chemistry and bond-breaking.
\newblock \emph{Journal of Chemical Theory and Computation}, 18\penalty0 (12):\penalty0 7024--7042, 2022.

\bibitem[Lewis et~al.(2020)Lewis, Perez, Piktus, Petroni, Karpukhin, Goyal, K{\"u}ttler, Lewis, Yih, Rockt{\"a}schel, et~al.]{lewis2020retrieval}
Patrick Lewis, Ethan Perez, Aleksandra Piktus, Fabio Petroni, Vladimir Karpukhin, Naman Goyal, Heinrich K{\"u}ttler, Mike Lewis, Wen-tau Yih, Tim Rockt{\"a}schel, et~al.
\newblock Retrieval-augmented generation for knowledge-intensive nlp tasks.
\newblock \emph{Advances in Neural Information Processing Systems}, 33:\penalty0 9459--9474, 2020.

\bibitem[Lewis et~al.(2021)Lewis, Wu, Liu, Minervini, K{\"u}ttler, Piktus, Stenetorp, and Riedel]{lewis2021paq}
Patrick Lewis, Yuxiang Wu, Linqing Liu, Pasquale Minervini, Heinrich K{\"u}ttler, Aleksandra Piktus, Pontus Stenetorp, and Sebastian Riedel.
\newblock Paq: 65 million probably-asked questions and what you can do with them.
\newblock \emph{Transactions of the Association for Computational Linguistics}, 9:\penalty0 1098--1115, 2021.

\bibitem[Li et~al.(2024)Li, Qin, Xiao, Chen, Luo, Shao, Lian, and Liu]{li2024makingtextembeddersfewshot}
Chaofan Li, MingHao Qin, Shitao Xiao, Jianlyu Chen, Kun Luo, Yingxia Shao, Defu Lian, and Zheng Liu.
\newblock Making text embedders few-shot learners, 2024.
\newblock URL \url{https://arxiv.org/abs/2409.15700}.

\bibitem[Liu et~al.(2018)Liu, Chen, Deng, Zeng, Chen, Li, and Tang]{liu2018lcqmc}
Xin Liu, Qingcai Chen, Chong Deng, Huajun Zeng, Jing Chen, Dongfang Li, and Buzhou Tang.
\newblock Lcqmc: A large-scale chinese question matching corpus.
\newblock In \emph{Proceedings of the 27th international conference on computational linguistics}, pages 1952--1962, 2018.

\bibitem[Liu et~al.(2024)Liu, Ping, Roy, Xu, Shoeybi, and Catanzaro]{liu2024chatqa}
Zihan Liu, Wei Ping, Rajarshi Roy, Peng Xu, Mohammad Shoeybi, and Bryan Catanzaro.
\newblock Chatqa: Building gpt-4 level conversational qa models.
\newblock \emph{arXiv preprint arXiv:2401.10225}, 2024.

\bibitem[Lo et~al.(2020)Lo, Wang, Neumann, Kinney, and Weld]{lo-wang-2020-s2orc}
Kyle Lo, Lucy~Lu Wang, Mark Neumann, Rodney Kinney, and Daniel Weld.
\newblock {S}2{ORC}: The semantic scholar open research corpus.
\newblock In \emph{Proceedings of the 58th Annual Meeting of the Association for Computational Linguistics}, pages 4969--4983, Online, July 2020. Association for Computational Linguistics.
\newblock \doi{10.18653/v1/2020.acl-main.447}.
\newblock URL \url{https://www.aclweb.org/anthology/2020.acl-main.447}.

\bibitem[Maia et~al.(2018)Maia, Handschuh, Freitas, Davis, McDermott, Zarrouk, and Balahur]{maia201818}
Macedo Maia, Siegfried Handschuh, Andr{\'e} Freitas, Brian Davis, Ross McDermott, Manel Zarrouk, and Alexandra Balahur.
\newblock Www'18 open challenge: financial opinion mining and question answering.
\newblock In \emph{Companion proceedings of the the web conference 2018}, pages 1941--1942, 2018.

\bibitem[Mecklenburg et~al.(2024)Mecklenburg, Lin, Li, Holstein, Nunes, Malvar, Silva, Chandra, Aski, Yannam, et~al.]{mecklenburg2024injecting}
Nick Mecklenburg, Yiyou Lin, Xiaoxiao Li, Daniel Holstein, Leonardo Nunes, Sara Malvar, Bruno Silva, Ranveer Chandra, Vijay Aski, Pavan Kumar~Reddy Yannam, et~al.
\newblock Injecting new knowledge into large language models via supervised fine-tuning.
\newblock \emph{arXiv preprint arXiv:2404.00213}, 2024.

\bibitem[Meng* et~al.(2024)Meng*, Liu*, Joty, Xiong, Zhou, and Yavuz]{SFR-embedding-2}
Rui Meng*, Ye~Liu*, Shafiq~Rayhan Joty, Caiming Xiong, Yingbo Zhou, and Semih Yavuz.
\newblock Sfr-embedding-2: Advanced text embedding with multi-stage training, 2024.
\newblock URL \url{https://huggingface.co/Salesforce/SFR-Embedding-2_R}.

\bibitem[Meng et~al.(2024)Meng, Liu, Joty, Xiong, Zhou, and Yavuz]{SFRAIResearch2024}
Rui Meng, Ye~Liu, Shafiq~Rayhan Joty, Caiming Xiong, Yingbo Zhou, and Semih Yavuz.
\newblock Sfr-embedding-mistral:enhance text retrieval with transfer learning.
\newblock Salesforce AI Research Blog, 2024.
\newblock URL \url{https://blog.salesforceairesearch.com/sfr-embedded-mistral/}.

\bibitem[Thakur et~al.(2021)Thakur, Reimers, R{\"u}ckl{\'e}, Srivastava, and Gurevych]{thakur2021beir}
Nandan Thakur, Nils Reimers, Andreas R{\"u}ckl{\'e}, Abhishek Srivastava, and Iryna Gurevych.
\newblock Beir: A heterogenous benchmark for zero-shot evaluation of information retrieval models.
\newblock \emph{arXiv preprint arXiv:2104.08663}, 2021.

\bibitem[Wadden et~al.(2020)Wadden, Lin, Lo, Wang, van Zuylen, Cohan, and Hajishirzi]{wadden-etal-2020-fact}
David Wadden, Shanchuan Lin, Kyle Lo, Lucy~Lu Wang, Madeleine van Zuylen, Arman Cohan, and Hannaneh Hajishirzi.
\newblock Fact or fiction: Verifying scientific claims.
\newblock In \emph{Proceedings of the 2020 Conference on Empirical Methods in Natural Language Processing (EMNLP)}, pages 7534--7550, Online, November 2020. Association for Computational Linguistics.
\newblock \doi{10.18653/v1/2020.emnlp-main.609}.
\newblock URL \url{https://aclanthology.org/2020.emnlp-main.609}.

\bibitem[Wang et~al.(2023)Wang, Yang, Huang, Yang, Majumder, and Wei]{wang2023improving}
Liang Wang, Nan Yang, Xiaolong Huang, Linjun Yang, Rangan Majumder, and Furu Wei.
\newblock Improving text embeddings with large language models.
\newblock \emph{arXiv preprint arXiv:2401.00368}, 2023.

\bibitem[Xie et~al.(2023)Xie, Dong, Wang, Lv, Yao, Gan, Wu, Li, Li, Liu, et~al.]{xie2023t2ranking}
Xiaohui Xie, Qian Dong, Bingning Wang, Feiyang Lv, Ting Yao, Weinan Gan, Zhijing Wu, Xiangsheng Li, Haitao Li, Yiqun Liu, et~al.
\newblock T2ranking: A large-scale chinese benchmark for passage ranking.
\newblock In \emph{Proceedings of the 46th International ACM SIGIR Conference on Research and Development in Information Retrieval}, pages 2681--2690, 2023.

\bibitem[Yang et~al.(2019)Yang, Zhang, Tar, and Baldridge]{yang2019paws}
Yinfei Yang, Yuan Zhang, Chris Tar, and Jason Baldridge.
\newblock Paws-x: A cross-lingual adversarial dataset for paraphrase identification.
\newblock \emph{arXiv preprint arXiv:1908.11828}, 2019.

\bibitem[Zhou et~al.(2022)Zhou, Gong, Liu, Zhao, Shen, Dong, Lu, Majumder, Wen, Duan, et~al.]{zhou2022simans}
Kun Zhou, Yeyun Gong, Xiao Liu, Wayne~Xin Zhao, Yelong Shen, Anlei Dong, Jingwen Lu, Rangan Majumder, Ji-Rong Wen, Nan Duan, et~al.
\newblock Simans: Simple ambiguous negatives sampling for dense text retrieval.
\newblock \emph{arXiv preprint arXiv:2210.11773}, 2022.

\end{thebibliography}



\end{document}